\documentclass[useAMS,usenatbib]{mn2e}

\usepackage{epsfig}
\usepackage{longtable}
\usepackage{times} 
\newcommand{\sw}{$Swift$}

\def \src {\mbox{IGR~J16195--4945}}

\def \sw {{\em Swift}}

\def \ferg {erg cm$^{-2}$ s$^{-1}$}
\def \hcm {\hbox {\ifmmode $ atom cm$^{-2}\else atom cm$^{-2}$\fi}}

\begin{document}

\title[The orbital period of IGR~J16195-4945]
{\emph{Swift} reveals the eclipsing nature of the high mass X-ray binary IGR~J16195-4945
}

\author[G.\ Cusumano et al.]{G.\ Cusumano$^{1}$, V.\ La Parola$^{1}$, 
A.\ Segreto $^{1}$, A.\ D'A\`i$^{1}$\\
$^{1}$INAF, Istituto di Astrofisica Spaziale e Fisica Cosmica,
        Via U.\ La Malfa 153, I-90146 Palermo, Italy\\
}

\date{}

\pagerange{\pageref{firstpage}--\pageref{lastpage}} \pubyear{2015}

\maketitle

\label{firstpage}

\begin{abstract}
\src\ is a hard X-ray source discovered by INTEGRAL during the Core Program 
observations performed in 2003.
We analyzed the X-ray emission of this source exploiting the {\emph Swift}-BAT 
survey data from December 2004 to March 2015, and all the available 
{\emph Swift}-XRT pointed observations.

The source is detected at a high significance level 
in the 123-month BAT survey data, with an average 15--150
keV flux of the source of $\sim 1.6$ mCrab.
The timing analysis on the BAT data reveals with a significance higher than 6
standard deviations the presence of a modulated signal 
with a period of 3.945 d, that we interpret as the orbital period of the 
binary system. The 
folded light curve shows a flat profile with a narrow full eclipse lasting 
$\sim 3.5\%$ of the orbital period. We requested phase-constrained XRT
observations to obtain a more detailed characterization of the eclipse in the
soft X-ray range. 
Adopting resonable guess values for 
the mass and radius of the companion star, we derive a semi-major
orbital axis of $\rm \sim 31 R_{\odot}$, equivalent to $\sim 1.8$ times the radius of the
companion star. From these estimates and from the duration of the eclipse we derive 
an orbital inclination between 55 and 60 degrees.

The broad band time-averaged XRT+BAT spectrum is well modeled with a 
strongly absorbed 
flat power law, with absorbing column $\rm N_{H}=7\times 10^{22} cm^{-2}$ and 
photon index $\Gamma=0.5$, modified by a high 
energy exponential cutoff at $E_{cut}=14$ keV.


\end{abstract}

\begin{keywords}
X-rays: binaries -- X-rays: individual: IGR~J16195$-$4945. 

\noindent
Facility: {\it Swift}

\end{keywords}


        \section{Introduction\label{intro}}

Thanks to the imaging capability of the Burst Alert Telescope (BAT, \citealp{bat}) on 
board of the \emph{Swift}  observatory \citep{swift}
we have an all-sky, nearly continuous in time and spatially resolved monitoring in 
the 15-150 keV energy band since December 2004.
BAT observes daily $\sim$90 per cent of the sky thanks to its very large 
field of view (1.4 steradian half coded) and to the pointing strategy of the 
\emph{Swift} observatory that performs tens of pointings  per day towards different 
directions of the sky. This huge data set has proved extremely useful for 
spectral and temporal studies of numerous galactic and extragalactic sources. 

In this paper we exploit the hard X-ray monitoring collected by BAT on 
IGR~J16195--4945.  We assess the binary nature of the source through the detection of its
orbital period and derive information on the geometry of the system. Moreover, 
broad band spectral analysis is performed using also the soft X-ray data collected 
by the X-ray Telescope (XRT, \citealp{Burrows2005}) during the pointed 
observations performed by \emph{Swift}.

IGR~J16195--4945 is a faint source discovered by INTEGRAL during the Core Program 
observations performed between 27 February and 19 October 2003 \citep{walter}.
The source was detected in the direction of the  Norma  region  of  the  Galaxy 
($332^{\circ}< l < 334^{\circ}$, at $\sim 5$ kpc \citealp{russeil}). In the 4th 
IBIS/ISGRI catalogue \citep{bird}  IGR~J16195--4945 is 
reported with an intensity of $\sim2.0$ and $\sim1.2$ mCrab in the energy ranges 
20--40 keV and 40--100 keV, respectively.

\citet{sidoli05} proposed the ASCA X-ray source AX J161929--4945  as the likely
soft X-ray 
counterpart of this source. The analysis of archival ASCA observations showed 
a variable flux  with a variation of the source intensity up to a factor of $\sim2.5$. 
The ASCA spectrum of the source was modeled by an absorbed power law 
with  a photon index of $\sim0.6$, a high value of $\rm N_H\sim12 \times 10^{22}$ 
cm$^{-2}$)  and an observed flux of $1.6 \times 10^{-11}$ \ferg in 2-10 keV. 
Analyzing archival INTEGRAL data 
\citet{sidoli05} detected IGR~J16195--4945 in only 2 observations out of
56 (2003 March 4 and 14) 
both with an average flux level of $\sim17$ mCrab in the 20--40 keV range.

During an INTEGRAL observation performed on 2003 September 26th the source
showed a flare lasting $\sim1.5$ hours reaching a peak flux of $\sim35$ mCrab in 20--40 keV 
\citep{sguera}. Because of the time scale of the event, the source was classified as a 
possible Supergiant Fast X-ray Transient (SFXT) candidate.

Chandra  observed  the  source  on  2005  April  29th  allowing to
refine its position  at RA= 16h 19m 32.20s, Dec=$-49^{\circ} 44' 30.7'' $
(J2000),   with  an accuracy of 0.6 arcsec \citep{tomsick06}. 
This in turn allowed  to find the nIR and mIR counterparts in the 
2MASS (2MASS J16193220-4944305) and in the GLIMPSE (G333.5571 + 00.3390) catalogs, 
respectively. The Chandra spectrum was best fitted by  an  absorbed power law 
(Gamma $\sim0.5$ and $\rm N_H\sim7 \times 10^{22}$ cm$^{-2}$, unabsorbed flux  of  
$4.6 \times 12^{-12}$ \ferg in the 0.3--10 keV range).
Extracting a time-averaged spectrum using all the available public INTEGRAL data from 2003 
February 27 to 2004 September 15 \citet{tomsick06} found a best-fitting powerlaw
with photon index $\sim1.7$ and a 20--50 kev flux of 
$19 \times 10^{-12}$ \ferg. Assuming that the source is in the  Norma-Cygnus arm 
of the Galaxy and thus at a distance of 5 kpc, the derived X-ray luminosities are 
$1.4 \times 10^{34}$ (d/5kpc)$^2$ erg s$^{-1}$ in the 0.3-10 keV range and 
$5.8 \times 10^{34}$ (d/5kpc)$^2$ erg s$^{-1}$ in the 20-50 keV range. 
The nature of the companion star was constrained through the analysis of 
photometric and spectroscopic 
data  in the optical and near-infrared band, that allowed the identification of
the spectral type of the companion star as an ON9.7Iab supergiant stars 
\citep{coleiro}.

IGR~J16318-4848 was observed by Suzaku in September 2006. During this observation 
IGR~J16195-4945 showed a short ($\sim$ 5000 s) bright flare ($\sim10 \times$ brighter 
than the prior emission level) adding evidence for its SFXT nature \citep{morris}. 
The emission during the flare was modeled by an absorbed power-law and a partial 
covering model with a photon index of $\sim1.8$ and an average flux 
of $4.8 \times10^{-11}$ \ferg in the 0.2--10 keV range.
If IGR~J16195-4945 is a SFXT and assuming the equatorial disk wind model 
\citep{sidoli07}, that interprets the observed bright flare as the 
transit  of  the  compact  object   into  an  equatorial
disk wind of the donor star, and combining Suzaku data with the long-term Swift-BAT 
dataset, \citet{morris} predicted for this source an orbital period of $\sim$16 days.

This paper is organized as follows. Section 2 describes the data reduction.
 Section 3 reports on the timing analysis. 
In Sect. 4 we describe the spectral analysis and
in Sect. 5 we briefly discuss our results. 

        \section{Data Reduction\label{data}}

\begin{figure}
\begin{center}
\centerline{\includegraphics[width=9cm,angle=0]{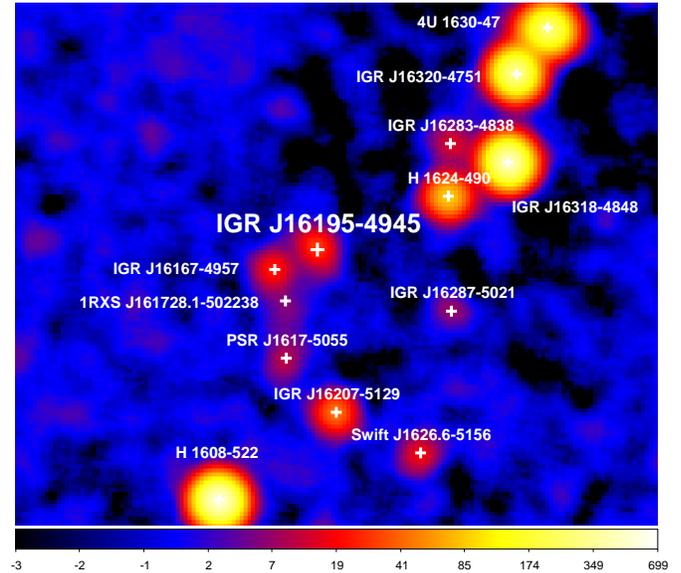}}
\caption[IGR~J16195--4945 sky maps]{ Left panel: 15--60 keV significance map in the 
neighborhood of IGR~J16195--4945.  
).
                }
               \label{map} 
        \end{center}
        \end{figure}

The BAT survey data collected between 2004 December and 2015 March  
were retrieved from the HEASARC public
archive\footnote{http://heasarc.gsfc.nasa.gov/docs/archive.html} and 
 processed with the  {\sc bat\_imager} code \citep{segreto}, a software built for 
the analysis of data from coded mask instruments that performs screening, 
mosaicking and source detection, and produces scientific products of any revealed 
source. 

IGR~J16195--4945 was detected in the 15--150 keV band at a significance of 28.5 
standard deviations, with a maximum of significance (31.7 standard deviations) in the 
15--60 keV energy band.
Figure~\ref{map} shows the 15--60 keV significance sky map (exposure time of 44.1 Ms) 
centered in the direction of IGR~J16195-4945.  
For the timing analysis of the BAT data we extracted the light curve in the 15--60 keV 
energy range with the maximum available time resolution of $\sim 300$ s 
while for 
the spectral analysis we produced the background subtracted spectrum of the source averaged over 
the entire exposure and we used the official BAT spectral redistribution
matrix\footnote{http://heasarc.gsfc.nasa.gov/docs/heasarc/caldb/data/swift/bat/index.html}

\begin{table*}
\begin{tabular}{r l l l l l l}
\hline
Obs \# & Obs ID          &$T_{start}$ &$T_{elapsed}$&Exposure (s) & Rate  & Orb. Phase  \\   
     &  		 &  (MJD)		 &  (s)    &  (s)  & (c/s)$\times10^{-2}$    &\\  \hline \hline
1    &    00036628001	 & 54385.749  &  104580  & 9055    & $12.4\pm 0.5 $ & 0.435--0.742	\\
2    &    00038000001	 & 54654.040  &   59057  & 7561    & $10.8\pm 0.5 $ & 0.442--0.616 \\
3    &    00038000002	 & 55350.107  &  29625   & 3646    & $1.4 \pm 0.3 $ & 0.885--0.972 \\
4    &    00042873001	 & 56079.489  &    469   &  469    & $33.0\pm 0.4 $ & 0.773--0.774 \\
5    &    00042866001	 & 56079.892  &    484   &  484    & $7.9 \pm 1.9 $ & 0.875--0.876 \\
6    &    00042866002	 & 56081.497  &    266   &  266    & $2.7 \pm 1.4 $ & 0.282--0.283 \\
7    &    00649143000	 & 57219.140  &   1321   & 1316    & $1.5 \pm 0.5 $ & 0.658--0.662 \\
8    &    00649143001	 & 57219.206  &   6260   & 2170    & $2.8 \pm 0.5 $ & 0.675--0.693 \\
9    &    00037887001	 & 57295.035  &    659   & 659     & $0.8 \pm 0.5 $ & 0.895--0.897 \\
10   &    00037887002	 & 57295.102  &    522   & 522     & $0.7 \pm 0.5 $ & 0.913--0.914 \\
11   &    00037887003	 & 57295.164  &    742   & 742     & $<1.48	  $ & 0.928--0.930 \\
12   &    00037887004	 & 57295.230  &    577   & 577     & $<3.24	  $ & 0.945--0.946 \\
13   &    00037887005	 & 57295.297  &    504   & 504     & $<3.05	  $ & 0.962--0.964 \\
14   &    00037887006	 & 57295.367  &    637   & 637     & $<1.78	  $ & 0.980--0.982 \\
15   &    00037887007	 & 57295.434  &    509   & 509     & $<2.24	  $ & 0.997--0.998 \\
16   &    00037887008	 & 57295.496  &    527   & 527     & $<2.15	  $ & 0.013--0.014 \\
17   &    00037887009	 & 57295.563  &    557   & 557     & $<2.05	  $ & 0.029--0.031 \\
18   &    00037887010	 & 57295.637  &    612   & 612     & $<1.85	  $ & 0.048--0.049 \\
19   &    00037887011	 & 57295.699  &    539   & 539     & $<3.48	  $ & 0.064--0.066 \\
20   &    00037887012	 & 57295.766  &    452   & 452     & $<3.40	  $ & 0.081--0.082 \\
21   &    00037887013	 & 57295.840  &    492   & 492     & $8.5 \pm 1.7 $ & 0.099--0.101 \\
22   &    00037887014	 & 57295.906  &    552   & 552     & $5.7 \pm 1.4 $ & 0.116--0.118 \\
23   &    00037887015	 & 57295.965  &    384   & 384     & $4.1 \pm 1.4 $ & 0.132--0.133 \\
\hline						 
\end{tabular}					 
\caption{Log of the Swift-XRT observations. \label{log}}
\end{table*}					 
						 
We have analysed all the available \sw-XRT observations of IGR~J16195--4945.
The source was always observed in Photon Counting (PC) mode.
In Table 1 we show the log of the
\sw-XRT observations and the relevant source count rates.
The XRT data were processed with standard procedures
({\sc xrtpipeline v.0.12.4}), filtering and screening criteria, using
ftools in the Heasoft package (v 6.12), adopting standard grade filtering 0-12.
The event arrival times were reported to the Solar System Baricenter using the
task {\sc
barycorr}\footnote{http://heasarc.gsfc.nasa.gov/ftools/caldb/help/barycorr.html}.
For each observation we extracted the source data from a circular region of 20 pixel radius (1
pixel = 2.36'') centered on the source position as determined with
{\sc xrtcentroid}\footnote{http://heasarc.gsfc.nasa.gov/ftools/caldb/help/xrtcentroid.html}.
 The spectra were re-binned
with a minimum of 20 counts per energy channel. This lower limit
of counts per bin is enough to ensure that the deviation of the observed 
number of counts from the expected values approximates
quite well a Gaussian distribution, that is a requirement to apply the
 $\chi^2$ statistics.
The background was extracted from an annular region
centered on the source with radii of 70 and 130 pixels.
XRT ancillary response files were generated with 
{\sc xrtmkarf}\footnote{http://heasarc.gsfc.nasa.gov/ftools/caldb/help/xrtmkarf.html}

We used the spectral redistribution matrix v014 and
the spectral analysis was performed using XSPEC v.12.5.
Errors are at 90\,\% confidence level, if not stated otherwise.

\begin{figure}
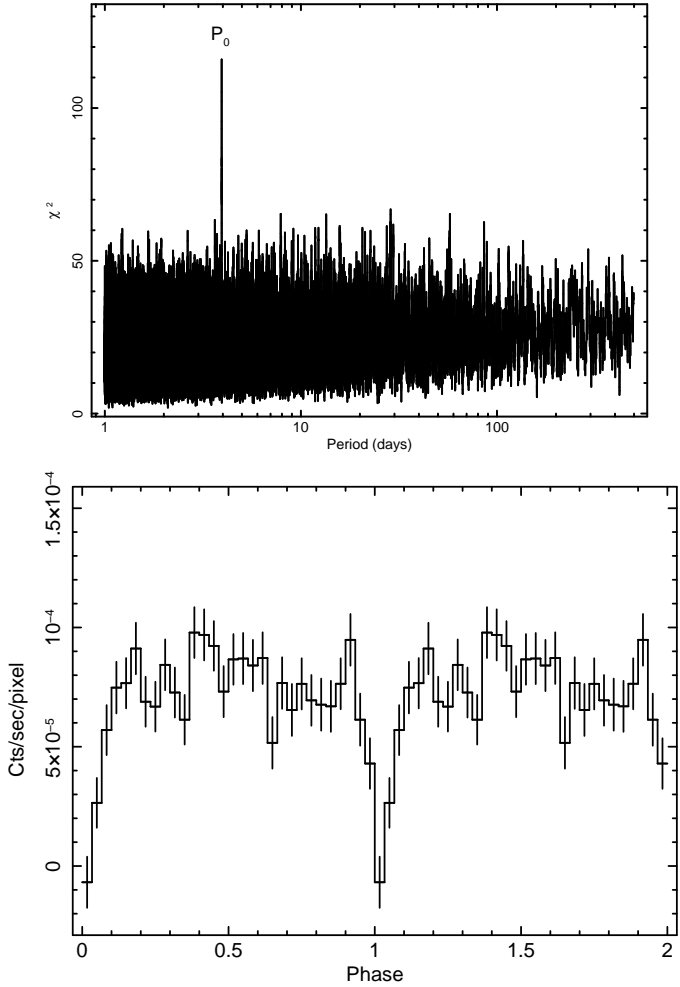

\begin{center}
\centerline{\includegraphics[width=6cm,angle=270]{figura2_bis.ps}}
\centerline{\includegraphics[width=7cm,angle=270]{figura_folding.ps}}

\caption[]{{\bf Top panel}: Periodogram of \sw/BAT (15--60\,keV) data for 
IGR~J16195--4945. 
{\bf Bottom panel}: Light curve folded at a period $P=  3.945$\,day, with 30 phase 
bins. 
}               
		\label{period} 
        \end{center}
        \end{figure}
	
        \section{Timing analysis\label{sfxt7:timing}}

We analyzed the long term BAT light curve searching for any periodic signal.
The 15--60 keV BAT light curve was folded with different trial periods P from  1.0\,d to 500\,d
with a step of $P^{2}/(N \,\Delta T)$, where $N=16$ is the number of trial profile 
phase bins and $\Delta T\sim$323 Ms is the data span length. To build the profile for each
trial period the average rate in each phase bin was evaluated
by weighting the rates with the inverse square of their statistical
error. This procedure is mandatory for data collected by a large
field of view coded mask telescope like BAT that is characterized
by a large spread of statistical count rate errors.
Figure~\ref{period} (Top panel) shows the resulting periodogram, where a single 
feature emerges, reaching a $\chi^2$ value of $\sim116$, at a period of 
P$_0=3.945\pm0.005$ d, where P$_0$ and its error are the centroid and the standard deviation 
obtained by modeling this feature with a Gaussian function.
We observe that in a wide interval of trial periods 
(1--50 d) around the feature position  the average $\chi^2$ is $\sim20$, and
therefore it deviates significantly from the average value of N-1 expected for 
the $\chi^2$ statistics for a white noise signal. As a consequence, to estimate
the significance of P$_0$ we applied the method described in \citet{segreto13},
obtaining that the probability of finding a value of $\chi^2$ equal or higher than
116 is $1.8\times10^{-10}$, corresponding to more than 6 standard deviations in
Gaussian statistics.

\begin{figure}
\begin{center}
\centerline{\includegraphics[width=9cm]{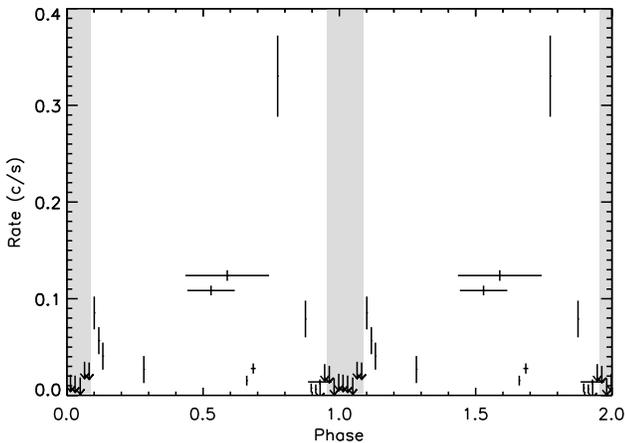}}

\caption[]{XRT count rate of the pointed observations vs orbital phase. The
shaded area is centered on the minimum in the BAT folded light curve and its
width is FWHM of the Gaussian that best fit the dip.
}               
		\label{xrtprof} 
        \end{center}
        \end{figure}

The BAT light curve folded at P$_0$ with T$_{\rm epoch}$=55220.4179 MJD shows a quite flat 
intensity level and a sharp dip consistent with no emission. 
To evaluate the phase position of the dip centroid we built a new profile,
adopting a finer (N=30) phase bin grid (Fig~\ref{period}, central 
panel) and fitted the dip 
with a Gaussian model: the centroid is at phase $0.02\pm 0.01$ corresponding to 
$(55220.50\pm0.04)\pm nP_0$ MJD, while the Full Width Half Maximum (FWHM) is 
$0.08\pm 0.02$ in phase, corresponding to $0.32\pm 0.07$ d.
To confirm  the presence of the dip in the soft X-ray band, we 
requested a \sw-XRT monitoring of the source in the phase
interval around the dip (Obs \# 9--23 in Table~\ref{log}, corresponding to
orbital phases 0.91--1.13). Figure~\ref{xrtprof}
shows the XRT count rates of all the XRT observations folded at P$_0$. 
The shaded area corresponds to the phase interval of the BAT dip. In all the
observations whithin the dip phase interval the source is not detected, strongly
suggesting an eclipse of the compact source by the companion star. Cumulating 
observations 11 to 20 we obtain a more stringent upper limit to the source count
rate of $2.4\times 10^{-3}$ count/s.

Where the source is detected, its count rate shows significant variability, with a variation of a
factor of $\sim30$ between the two observations with extreme rate values (Obs 4 and
10). An inspection of the light curves extracted from each observation shows a
rate variation of a factor of $< 5$ within each dataset.

We performed a search for periodic modulations only on the XRT data from 
observations 1 and 2,
as the statistics of the other datasets is too low for this kind of analysis.
In order to avoid systematics caused by the read-out time in PC mode 
characterized by a time resolution bin of $\delta T_{XRT}$=2.5073 s, 
the arrival times of the events in PC mode were randomized within 
$\delta T_{XRT}$. 
We performed a folding analysis on the source data from each observation 
searching in the period range between $\delta T_{XRT}$ and 1000 s.
We did not find any significant feature in the resulting periodograms.

\section{Broad band spectral analysis\label{spetra}}

We performed a broad band spectral analysis  in the 0.2-150 keV energy range.
We used only the XRT data extracted from observations
1 and 2; the other XRT observations were not used because of their low 
statistics. A preliminary analysis showed no significant differences between the two 
spectra, and they were therefore summed into a single spectrum. 
We also checked for the presence of any significant spectral variability during the BAT monitoring 
through the inspection of the hardness ratio calculated in the energy ranges [35--85 keV][15--35 keV]
with several time bins between 1 and 20 days. We found no significant variability on any time scale.

Therefore we performed the broad band spectral analysis coupling the soft X-ray spectrum with the BAT 
spectrum averaged over 123 months, adding a constant factor to take into account both an 
intercalibration factor between the two telescopes and the different flux level between the two 
datasets. An absorbed power law model gave an unacceptable $\chi^2$ of 165 
with 76 d.o.f., with evident residuals between the data and the best 
fit model. The spectra indeed resulted well fitted ($\chi^2$=80.0 with 75 
d.o.f.) using a power law with a cutoff at $\sim 14$ keV absorbed by a column
of $\rm N_H=(7\pm1)\times 10^{22} cm^{-2}$ (model {\tt tbabs*(cutoffpl)} in 
{\sc xspec}).

Figure~\ref{spec} shows the combined XRT and BAT spectral data 
with the best fit model (top panel) and the residuals in unit of standard deviations (bottom panel).
Table~\ref{fit} reports the best fit parameters.

\section{Discussion\label{discuss}}
In this paper we have exploited the {\it Swift} archival data set on
IGR~J16195-4945. The source is detected at a level of 28.5 standard deviations
cumulating the first 123 months of BAT survey data, with an average flux of 
$(2.2\pm 0.2) \times 10^{-11}$ erg s$^{-1}$ cm$^{-2}$ in the 15-150 keV band,
corresponding to $(6.5 \pm 0.6)\times 10^{34}$ erg s$^{-1}$ assuming a distance of 5 kpc.
The intrinsic flux in the soft X-ray band (0.2-10 keV) ranges between $0.3\times
10^{-11}$ erg s$^{-1}$ cm$^{-2}$ and $6.2\times10^{-11}$ erg cm$^{-2}$ s$^{-1}$,
corresponding to a luminosity of $8.9\times10^{33}$ erg s$^{-1}$ and
$1.8\times10^{35}$ erg s$^{-1}$, respectively.
We have performed a broad band 15-150 keV spectral analysis combining the XRT 
and BAT average spectra. The source spectrum is well modeled with an absorbed
power-law with a cutoff at $\sim 14$ keV. The column density and the power-law
slope are $\sim 7\times10^{22}$ cm$^{-2}$ and $\sim 0.5$, respectively, in full 
agreement with the values observed with Chandra \citep{tomsick06}. This spectral 
shape is typical for high mass X-ray binary pulsars \citep{coburn02}.

The timing analysis performed on the BAT light curve reveals the presence of a
modulated signal with a periodicity P$_0=3.945\pm0.005$ d, that we interpret as
the orbital period of this binary system. Using indicatively the same mass and 
radius derived for the O9.7-type companion star in the Cyg X-1 system   
(M$_{\star}\sim 24$ M$_{\odot}$ and R$_{\star}\sim 17$ R$_{\odot}$, 
respectively, \citealp{ziolkowski14}), 
we can derive the semi-major axis of the system through the application of the 
Kepler's third law:
\begin{equation}
a=(G P_o^2~(M_{\star}+M_{\rm X})/4\pi^2)^{1/3} \simeq 31 R_{\odot}  \simeq 1.8 R_{\star}
\end{equation}
if we assume the compact object to be a neutron star with 
$M_{\rm X}=1.4 M_{\odot}$. 
The BAT light curve folded with a period of P$_0$ shows a flat profile
interrupted by a full eclipse with estimated mid-eclipse time  of MJD 
$\rm (55220.50\pm0.04)\pm nP_0$. The monitoring of \sw-XRT shows that the source
is not detected during the eclipse phase interval, with a $3\sigma$ 
upper limit of $4.4\times10^{-13}$ \ferg, while it is always well 
visible outside this interval.
The full eclipse lasts $\sim 3.5\%$ of the orbital period ($\sim0.14$ d): such a short
duration corresponds to an inclination of the orbit between 55 and 60
degrees, under the hypothesis that the orbit has a very low eccentricity,
as the flatness of the folded light curve suggests.

\begin{figure}
\begin{center}
\centerline{\includegraphics[width=6cm,angle=-90]{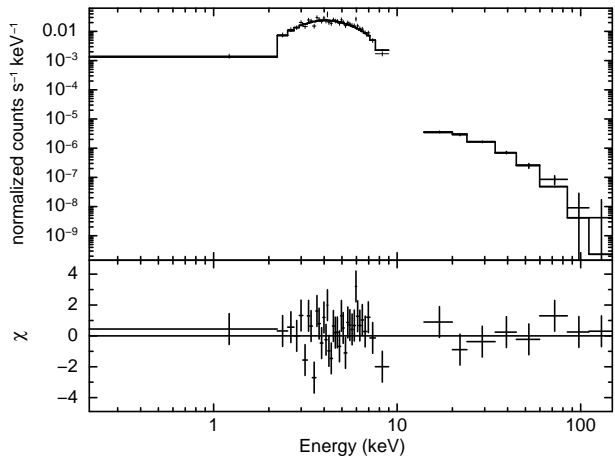}}
\caption[IGR J16195-4945 XRT and BAT spectrum]{{\bf Top panel}:
IGR J16195-4945 spectrum (XRT + BAT) and best fit model. 
{\bf Bottom panel}: Residuals in unit of standard deviations.
}
                \label{spec} 
        \end{center}
        \end{figure}

We have analyzed the long term {\it Swift}-BAT light curve of the source
IGR~J16195-4945 to study its variability. The source light curve in the 15-60
keV band (where the source has the highest significance) was binned into
1h, 2h, 6h, 12h intervals in order to examine the source behavior on these
timescales. The average intensity level of the source is $\sim 1.6$ mCrab. The
most significant deviations from this average level (with a S/N ratio of 
$> 5$) reach an intensity of $\sim 50$ mCrab on hours time scale, and
$\sim 15$ mCrab with a time bin of 12h. However, the overall distribution of the 
observed S/N values in the relevant light curves shows that the significance 
values of these peaks are only marginal outliers with respect to the observed
fluctuations. We conclude that short time flares of the order of tenth of mCrab
(such as those observed with other satellites, \citealp{sguera,morris})
cannot be revealed with the sensitivity of the BAT light curves.
On the other hand, the pointed XRT observations outside the eclipse, 
although did not show any clear flaring episode within a single observation,
differ in the observed averaged count rates up to a factor of 20.
Assuming the best fit broad-band spectral 
model (Table~\ref{spec}), the
two extreme XRT count rate values, extrapolated to the 15-60 keV band, 
correspond to 0.36 and 7.6 mCrab, respectively.

\begin{table}
\begin{tabular}{ r l l}
\hline
Parameter & Best fit value & Units    \\ \hline \hline
$\rm N_H$    & $7.0^{+1.3}_{-1.1}\times 10^{22}$ & cm$^{-2}$\\
$\Gamma$  &$0.5^{+0.3}_{-0.3}$&          \\
$E_{cut}$ &$14^{+3}_{-2}$ & keV\\
$N$       &$ (1.3^{+0.9}_{-0.5})\times 10^{-3}$ &ph $\rm keV^{-1} cm^{-2} s^{-1}$ at 1 keV \\
$\rm C_{BAT}$&$0.49^{+0.14}_{-0.11}$&\\
$\rm F$ (0.2--10 keV)&$(2.7\pm 0.3) \times 10^{-11}$& erg s$^{-1}$ cm$^{-2}$\\
$\rm F$ (15--150 keV)&$(2.2\pm 0.2 \times 10^{-11}$& erg s$^{-1}$ cm$^{-2}$\\
$\chi^2$   &80.1 (75 d.o.f.) & \\ \hline
\end{tabular}
\caption{Best fit spectral parameters.  We report
unabsorbed fluxes for the characteristic XRT (0.2--10 keV) and BAT (15--150 keV)
energy bands. \label{fit}}
\end{table}

\section*{Acknowledgments}

This work was supported in Italy by ASI contract I/004/11/1.
We thank the \sw~ team for having promptly performed the ToO observations
analysed in this paper.

\bsp

\label{lastpage}

\end{document}